# Unique information from common diffusion MRI models about white-matter differences across the human adult lifespan

Rafael Neto Henriques[1,*], Richard Henson[2,3], Cam-CAN[4] & Marta Morgado Correia[2]

[1] Champalimaud Research, Champalimaud Foundation, Portugal.

[2] MRC Cognition and Brain Sciences Unit, University of Cambridge, United Kingdom.

[3] Department of Psychiatry, University of Cambridge, United Kingdom.

[4] Cambridge Centre for Ageing and Neuroscience (Cam-CAN), University of Cambridge, United Kingdom

* corresponding author: rafael.henriques@neuro.fchampalimaud.org

# Acknowledgments

R.N.H. was supported by the Scientific Employment Stimulus 4th Edition from Fundação para a Ciência e Tecnologia, Portugal, ref 2021.02777.CEECIND. R.H. was supported by MRC programme grant SUAG/046 G101400. Cam-CAN was supported by the Biotechnology and Biological Sciences Research Council under grant BB/H008217/1. M.M.C. was supported by MRC Unit grant SUAG/019 G116768. For the purpose of open access, the author has applied a Creative Commons Attribution (CC BY) licence to any Author Accepted Manuscript version arising from this submission.

# Abstract

Diffusion Magnetic Resonance Imaging (dMRI) is sensitive to white matter microstructural changes across the human lifespan. Several models have been proposed to provide more sensitive and specific metrics than those provided by the conventional Diffusion Tensor Imaging (DTI) analysis. However, previous results using different metrics have led to contradictory conclusions regarding the effect of age on fibre demyelination and axonal loss in adults. Moreover, it remains unclear whether these metrics provide distinct information about the effects of age, e.g, on different white-matter tracts. To address this, we analysed dMRI data from 651 adults approximately uniformly aged from 18 to 88 years in the Cam-CAN cohort, using six dMRI metrics: Fractional Anisotropy (FA) from standard DTI; Mean Signal Diffusion (MSD) and Mean Signal Kurtosis (MSK) from Diffusional Kurtosis Imaging (DKI); and Neurite Density Index (NDI), Orientation Dispersion Index (ODI) and isotropic Free water volume fraction ($F_{iso}$) estimated from Neurite Orientation Dispersion and Density Imaging (NODDI). Averaging across white-matter regions-of-interest (ROIs), second-order polynomial fits revealed that MSD, MSK and $F_{iso}$ showed the strongest effects of age, with significant quadratic components suggesting more rapid and sometimes inverted effects in old age. Analysing the data in different age subgroups revealed that some apparent discrepancies in previous studies may be explained by the use of cohorts with different age ranges. Factor analysis of the six metrics across all ROIs revealed three independent factors that can be associated to 1) tissue non-Gaussian diffusion effects, 2) free-water contamination, and 3) tissue configuration complexity (e.g. crossing, dispersing, fanning fibres). While FA captures a combination of different factors, other dMRI metrics are strongly aligned to specific factors (NDI and MSK with Factor 1, $F_{iso}$ with Factor 2, and ODI with Factor 3). In summary, our study offers an explanation for previous discrepancies reported in dMRI ageing studies and provides further insights on the interpretation of different dMRI metrics in the context of white matter microstructural properties.

# 1. Introduction

Brain structure is known to change with age at many spatial scales. For instance, studies using conventional structural imaging techniques have shown that the volume of White Matter (WM) in many brain regions significantly decreases after the fifth decade of life (e.g., Bethlehem et al., 2022; Lebel et al., 2012; Walhovd et al., 2011). These gross morphological changes are likely to be a consequence of earlier microstructural alterations, such as cell loss, fibre loss, demyelination and increases of extra-cellular space (Aboitiz et al., 1996; Geula and Mesulam, 1989; Meier-Ruge et al., 1992; Pyapali and Turner, 1996; Scheltens et al., 1995), which are beyond the direct resolution limit of current magnetic resonance imaging (MRI). Fortunately, MRI can also be used to measure water diffusion in vivo – a modality known as diffusion MRI (dMRI). Since diffusion results in displacements of water molecules at a micrometric scale (during the timescales of typical MRI), dMRI can provide information at a scale below the dimension of its image voxels (Le Bihan and Johansen-Berg, 2012; Moseley, 2002). Previous studies have indeed shown that microstructural alterations measured by dMRI could occur before changes are observed on more conventional structural MRI contrasts (Maillard et al., 2013; Nusbaum et al., 2001; Pelletier et al., 2017).

The information captured by dMRI maps is multi-dimensional, and several models have been proposed to quantify different properties from dMRI images. Phenomenological dMRI models such as diffusion tensor imaging (DTI) and diffusional kurtosis imaging (DKI) can be used to summarise diffusion properties that can be indirectly related to properties of tissue microstructure (Basser et al., 1994; Jensen et al., 2005; Jensen and Helpern, 2010). Early human ageing studies using DTI showed that the anisotropy of diffusion in brain WM starts declining after the first two decades of life (Davis et al., 2009; Lebel et al., 2012; Pfefferbaum et al., 2000; Sullivan et al., 2001; Yeatman et al., 2014; Zhang et al., 2010). These initial declines were assumed to be associated with degenerative processes such as fibre demyelination and axonal loss. Studies using DKI showed that the degree of non-Gaussian diffusion increases up to the fifth decade of life (Coutu et al., 2014; Das et al., 2017; Falangola et al., 2008; Gong et al., 2014; Lätt et al., 2013). Since increased degree of non-Gaussian diffusion has been associated with WM maturation processes (Helpern et al., 2011a; Jensen and Helpern, 2010; Paydar et al., 2014), these DKI results are difficult to reconcile with the degeneration suggested by DTI anisotropy metrics. However, since DKI relies on subtle information from the non-linear behaviour of the log diffusion signal decay, the age-related profile provided by standard DKI metrics can be highly corrupted by thermal noise (Billiet et al., 2015; Henriques et al., 2021b; Tax et al., 2015; Veraart et al., 2011). Moreover, like any other phenomenological models, the interpretation of differences in DKI metrics is limited since they do not provide a direct link to specific microstructural properties.

Several more neuroanatomically-inspired models (also referred to as "mechanistic" or "microstructural" models, Novikov et al., 2018) have been proposed as an attempt to directly estimate specific tissue properties from diffusion-weighted images (e.g., Assaf et al., 2004; Assaf and Basser, 2005; Fieremans et al., 2011; Huber et al., 2019; Jespersen et al., 2007; Rokem et al., 2015; White et al., 2013). One of the most popular microstructural models used in clinical research is the "Neurite Orientation Dispersion and Density Imaging" (NODDI) model (Zhang et al., 2012). This has been used to estimate the degree of fibre dispersion (the "Orientation Dispersion Index", ODI) and neurite density (the "Neurite Density Index", NDI) in the context of ageing (e.g., Billiet et al., 2015; Chang et al., 2015; Kodiweera et al., 2016). These studies generally showed that early declines in diffusion anisotropy are most likely due to decrease of fibre dispersion (as measured by ODI). However, these studies also produced some inconsistencies. For example, while positive correlations between NDI and age were reported in some studies (Billiet et al., 2015; Chang et al., 2015), supporting the previous late maturation processes measured by DKI, negative correlations were reported by others (Cox et al., 2016; Merluzzi et al., 2016). These discrepancies may be a consequence of the low number of participants and/or variable age ranges used across these studies.

In an attempt to address this issue, Beck and colleagues compared a number of different dMRI techniques on a larger cohort of subjects covering the adult lifespan (18-94 years old), and found that the rates of change in DKI and NODDI metrics depended on age (Beck et al., 2021). However, their dMRI metrics were compared in terms of their average value across a whole-brain WM skeleton. The question of whether various metrics provide complementary information might vary across different WM tracts, depending on, for example, their degree of crossing fibres or proximity to ventricles. For example, the corona radiata have many crossing fibres, whereas the corpus callosum does not; and tracts such us the fornix will be more affected by free-water contamination. Furthermore, the question of whether dMRI metrics provide complementary information can also be addressed formally by principal component analysis (PCA). For example, Chamberland and colleagues found that only two principal components were necessary to capture most of the covariance between 10 dMRI metrics in a developmental dataset of 36 people aged 8-18 (Chamberland et al., 2019). Their first component captured properties related to hindrance and restriction in tissue microstructure (i.e., non-Gaussian diffusion effects), while their second component captured properties related to tissue configuration complexity (i.e., fibre crossing and dispersion effects). Here we perform a similar PCA, followed by axes rotation (i.e., Factor Analysis), of six different dMRI metrics, but now on a much larger, adult sample.

In summary, we compared the sensitivity to age of the main metrics from phenomenological and microstructural models of dMRI, using a large and homogeneous sample across the adult lifespan, namely from the Cambridge Centre for Ageing and Neuroscience (Cam-CAN) cohort (Shafto et al., 2014; Taylor et al., 2017). Stage 2 of this cohort includes dMRI data from 651 participants aged approximately uniformly from 18 to 88 years. Because these individuals were recruited via local doctors using an opt-out procedure, they are likely to be more representative of the effects of age than studies that recruit via advertisement (e.g., which tend to recruit super-healthy older people). As well as potentially resolving the inconsistencies in previous studies of ageing, we examined how six dMRI metrics vary across different sub-groups of age, across WM tracts, and also how they covaried across individuals in terms of underlying factors.

## 2. Methods

### 2.1 Data Acquisition

The data repository of Cam-CAN contains 651 complete diffusion-weighted datasets for healthy participants (319 males/332 females) with ages between 18 and 88 years (Taylor et al., 2017). These healthy participants were selected from 2681 interviewed participants with no serious psychiatric problems (Shafto et al., 2014; Taylor et al., 2017). Diffusion-weighted datasets were acquired on a 3T Siemens Trio Scanner (32-channel head coil) for two non-zero b-values (1000 and 2000 s/mm$^2$) along 30 diffusion gradient directions and for three b=0 volumes. A twice refocused spin echo (TRSE) echo-planar imaging sequence was used for eddy-current reduction (Reese et al., 2003). Other acquisition parameters were as follows: 66 axial slices, voxel size = 2×2×2 mm, TR= 9100 ms, TE= 104 ms, matrix = 96×96, field of view (FOV) = 192×192 mm$^2$, partial Fourier of 7/8, and acceleration factor of 2 using GRAPPA with 36 reference lines. More information about diffusion MRI acquisitions is reported in (Taylor et al., 2017). The raw data are in BIDS format are available on request from this website: https://camcan-archive.mrc-cbu.cam.ac.uk/dataaccess/. The code for all diffusion-weighted data pre-processing steps (vide infra) are available on the following repository: https://github.com/RafaelNH/CamCAN-dMRI-study.

### 2.2 Data Quality Control

The quality of the diffusion-weighted datasets was first visually inspected (Tournier et al., 2011). Based on this, two datasets were excluded: one because no anatomical information was acquired due to an acquisition failure and another because of abnormal cerebral ventricle sizes. In addition, 11 datasets were excluded because they possessed more than four volumes of diffusion-weighted images that were corrupted by motion-induced artefacts (i.e., image slice signal loss and "striping" pattern artefacts induced by motion during the acquisition of a single diffusion-weighted image, Tournier et al., 2011). The number of diffusion-weighted volumes corrupted by motion-induced artifacts was quantified using the procedure described in Supplementary Material Appendix A. A summary of the total number of included and excluded datasets for different participant age sub-groups is presented in Table 1.

Table 1 – Number of included and excluded diffusion-weighted datasets for different participant age sub-groups

| AGE (YEARS) | | DECILE 1 (18-27) | DECILE 2 (28-37) | DECILE 3 (38-47) | DECILE 4 (48-57) | DECILE 5 (58-67) | DECILE 6 (68-77) | DECILE 7 (78-88) | TOTAL |
|---|---|---|---|---|---|---|---|---|---|
| INCL. DATA | M | 22 | 50 | 43 | 51 | 47 | 46 | 50 | 309 |
| | F | 27 | 56 | 51 | 48 | 50 | 52 | 43 | 327 |
| | Total | 49 | 106 | 94 | 99 | 97 | 98 | 93 | 636 |
| EXCL. DATA | M | 2 | 0 | 1 | 0 | 1 | 3 | 3 | 10 |
| | F | 1 | 0 | 0 | 0 | 1 | 0 | 1 | 3 |
| | Total | 3 | 0 | 1 | 0 | 2 | 3 | 4 | 13 |

## 2.3 Data Pre-processing

Diffusion-weighted data was first denoised using a PCA based algorithm (Veraart et al., 2016) and then corrected for Gibbs artefacts using a sub-voxel shift procedure (Kellner et al., 2016). Then, data and respective gradient directions were corrected for motion misalignments using an adapted version of a procedure designed for high b-value diffusion-weighted images (Ben-Amitay et al., 2012), details of which are described in Supplementary Material Appendix B. After motion correction, non-brain voxels of processed datasets were removed using the brain extraction procedure of the FSL toolbox (Jenkinson et al., 2012; Smith, 2002).

## 2.4 Diffusion MRI Techniques

We focused on three dMRI models: 1) Diffusion Tensor Imaging (DTI) – the conventional, phenomenological dMRI technique; 2) Diffusional Kurtosis Imaging (DKI) – the next most used phenomenological model beyond DTI; and 3) the Neurite Orientation Dispersion and Density Imaging (NODDI) – the most common microstructural model. All models were fit in the native space of each participant (to decrease image artefact propagation due to data interpolation). Details of each dMRI model are reported below.

<u>Diffusion Tensor Imaging</u> – DTI was estimated using a non-linear, least-square solution (Jones and Basser, 2004; Koay et al., 2006) implemented on the open-source software package Diffusion written in Python (Garyfallidis et al., 2014; Henriques et al., 2021a). Only the conventional fractional anisotropy (FA) metric was estimated from the tensor; mean diffusivity (MD) was extracted using the DKI model - see below - because this decouples MD from higher-order kurtosis terms (Henriques et al., 2021a; Veraart et al., 2011). Other diffusion metrics, such as the axial and radial diffusivities, were not analysed, because their interpretation is challenging in WM regions containing complex crossing, fanning, or dispersing geometries (De Santis et al., 2014; Wheeler-Kingshott et al., 2009).

<u>Diffusional Kurtosis Imaging</u> – in this study, we focus on two directionally-averaged DKI estimates that are invariant to different WM configurations (i.e. invariant to presence of crossing fibres or to the degree of fibre dispersion and fanning). We are not interested in studying directional kurtosis quantities (e.g. axial and radial kurtosis) because their interpretation may be limited to aligned single WM fibre populations (Henriques et al., 2015), so a full diffusion and kurtosis tensor fitting was not performed. Instead, mean signal diffusion (MSD) and mean signal kurtosis (MSK) were directly extracted from averaged signals across different gradient directions (Henriques et al., 2021a, 2021b, 2019):

$$\log \bar{S}(b)/S_0 = -bMSD + \frac{1}{6}b^2 MSD^2 MSK + O(b^3) \qquad (1)$$

where $\bar{S}(b)$ represents the mean diffusion-weighted signals (signals averaged along different diffusion gradient directions for each individual b-value separately), and $S_0$ represents the mean signal for b-value=0. While MSD is equivalent to the standard MD computed from DKI, MSK provides similar results to the standard mean kurtosis index; however, mean signal estimates have the advantage of being more robust to thermal noise effects and invariant to different WM fibre configurations (Henriques et al., 2021a, 2019). For the present study, Equation 1 was fit using the weighted linear least-squares approach described by Henriques et al. (2021a).

<u>Neurite Orientation Dispersion and Density Imaging</u> – the NODDI model is a three-compartment model that was designed to estimate the NDI and ODI, while constraining all compartments' diffusivities to fixed priors to ensure model fit stability (Zhang et al., 2012). NODDI's model can be written as:

$$S(\boldsymbol{n}, b)/S_0 = (1 - F_{iso})[F_{ia}E_{ia}(\boldsymbol{n}, b) + (1 - F_{ia})E_{ea}(\boldsymbol{n}, b)] + F_{iso}E_{iso}(b) \quad (2)$$

with $F_{ia}$ being the intra-axonal volume fraction (i.e. $F_{ia}$ = NDI), $E_{ia}$ the intra-axonal signal attenuation, and $E_{ea}$ the extra-axonal signal attenuation. Note that this model also considered a third compartment to capture effects of isotropic diffusion of free water, with $F_{iso}$ and $E_{iso}$ representing its apparent volume fraction and signal attenuation. The signal attenuations for each compartment are given by:

$$E_{ia}(\boldsymbol{n}, b) = \int f(\boldsymbol{u}) \exp[-bd_\parallel (\boldsymbol{n}^T\boldsymbol{u})^2] \, d\Omega_{\boldsymbol{u}} \quad (3)$$

$$E_{ea}(\boldsymbol{n}, b) = \exp[-b\boldsymbol{n}^T (\int f(\boldsymbol{u}) D_e(\boldsymbol{u}) d\Omega_{\boldsymbol{u}})\boldsymbol{n}] \quad (4)$$

$$E_{iso}(b) = \exp(-bD_{iso}) \quad (5)$$

where $d_\parallel$ is the intrinsic axonal diffusivity, set to 1.7 µm²/ms; $D_e(\boldsymbol{u})$ is an axial symmetric tensor parallel to vector $\boldsymbol{u}$, with axial and radial diffusivities equal to $d_\parallel$ and $d_\perp = d_\parallel(1 - f_{ia})$; $D_{iso}$ is the isotropic free water diffusivity at the body temperature of 37°C, set to 3 µm²/ms; $f$ is the fibre orientation distribution function, which is assumed to follow a Watson distribution $f(\boldsymbol{n}) = {}_1F_1\left(\frac{1}{2}, \frac{3}{2}, k\right)^{-1} \exp[-k(\boldsymbol{\mu}^T\boldsymbol{n})^2]$, where ${}_1F_1$ is the confluent hypergeometric function of the first kind; $\boldsymbol{\mu}$ is the fibre average direction and $k$ is a metric related to ODI ($ODI = 2\arctan(1/k)/\pi$; Jespersen et al., 2012). Here NODDI was fit using the original implementation available at: http://www.nitrc.org/projects/noddi_toolbox (NODDI toolbox version 0.9).

In summary, we compared six dMRI metrics: 1) FA (from DTI); 2-3) MSD and MSK (from DKI); and 4-6) ODI, NDI, and $F_{iso}$ (from NODDI).

## 2.5 Data Analysis

The diffusion metric values for each participant and metric are available in the CSV files "Global_Metrics.csv" (averaged across WM voxels) and "ROI_Metrics.csv" (separately for each ROI) here: https://github.com/RafaelNH/CamCAN-dMRI-study. Matlab code for the statistical analysis can also be found there, in "main_dMRI_stats_analysis.m".

<u>Region of Interest (ROI) definition:</u> The values of diffusion-based metrics were averaged across voxels for each of the 48 WM ROIs included in the Johns Hopkins University (JHU) atlas (Mori et al., 2008). For this purpose, WM ROIs were warped from an FA template to each native FA map, using FSL's linear and non-linear registration tools (Jenkinson et al., 2012; Smith et al., 2004; Woolrich et al., 2009). To suppress the impact of cerebral spinal fluid (CSF) free-water partial volume effects (and to minimise the impact of degenerative ODI and NDI estimates on voxels containing most free water), voxels with $F_{iso}$ values larger than 0.9 were removed from the ROIs.

<u>Global WM analysis:</u> for global analysis, we averaged the dMRI values across all voxels in all ROIs. We then fit a second-order polynomial expansion of age (i.e., linear and quadratic terms), together with covariates of sex and the interaction between sex and polynomial age terms. For this analysis, we removed five participants whose residuals from this model were more than 5 standard deviations from the mean in at least one of the six dMRI metrics. Their ages were 30, 44, 45, 78 and 85 (i.e., not particularly biased to certain ages). We then refit the polynomial model and reported the proportion of variance explained by each effect.

ROI-specific analysis: to reduce the number of comparisons, we averaged data across those pairs of ROIs that were homologous across hemispheres, leading to a total of 27 ROIs remaining. We then Z-scored the values across participants for each ROI and metric. For each metric, we examined the distribution of resulting values (concatenated across ROI and participants) and removed participants whose data included a value more than 5 standard deviations from the mean. This was done to minimise the influence of extreme values on the PCA below, and resulted in removal of 20 participants, who tended to be either younger or older than the median age (six were 45 or under, and 14 were 68 or over), i.e, unlikely to systematically bias subsequent analyses towards young or older groups. This left 618 participants. The proportion of variance ($R^2$) explained by the linear and quadratic terms of a second-order polynomial expansion of age was calculated for each ROI and each metric, and then the ROIs ranked by this proportion.

Age correlations for each ROI in different age subgroups: This analysis was performed to assess the dependency of (linear) correlations between each metric and age across different age ranges. For this, correlations were calculated for three different age sub-groups with similar number of participants (approximately 200): participants aged 1) from 28 to 47 years; 2) from 48 to 67 years; and 3) from 68 to 87 years. To decrease the number of false positives, the false discovery rate (FDR) for the resulting 3 x 6 x 27 = 486 tests was controlled at q=0.05 (Benjamini and Hochberg, 1995).

Correlation between Metrics and Factor Analysis: The six metrics were concatenated across participant and ROIs (i.e., 15,450 observations per metric) and the Pearson correlation between each pair of them calculated before and after regressing out linear and quadratic effects of age. Principal Component Analysis (PCA) was then applied to the same matrix. Three PCs captured over 96% of the variance (see Results). Factor analysis was then applied by rotating three orthogonal axes to maximise the squared loadings ("Varimax").

# 3. Results

## 3.1 Representative dMRI Maps

For a qualitative inspection of the quality of the different diffusion MRI metrics, representative maps of the six diffusion MRI metrics considered in this study (FA, MSD, MSK, NDI, ODI, and $F_{iso}$) are shown in Fig. 1 for two young adults (26 and 25 years old, panels A and B) and for two elders (79 years old, panels C and D). In general, all diffusion metrics show the contrasts expected from previous literature (e.g. WM regions characterized by higher values for FA, MSK and NDI, and lower values of ODI, when compared to grey matter). MSK estimates in WM do not reveal the implausible negative kurtosis estimates reported in previous literature (e.g., Henriques et al., 2021b; Tabesh et al., 2011). Diffusion MRI maps for elders show enlarged ventricles (as highlighted by wider areas of MSD ≈ 3 $\mu m^2$/ms and wider areas of $F_{iso}$ ≈ 1 in panels C and D) and thinner WM fibre bundles (as revealed by the narrow WM areas in FA, MSK and NDI maps in panels C and D).

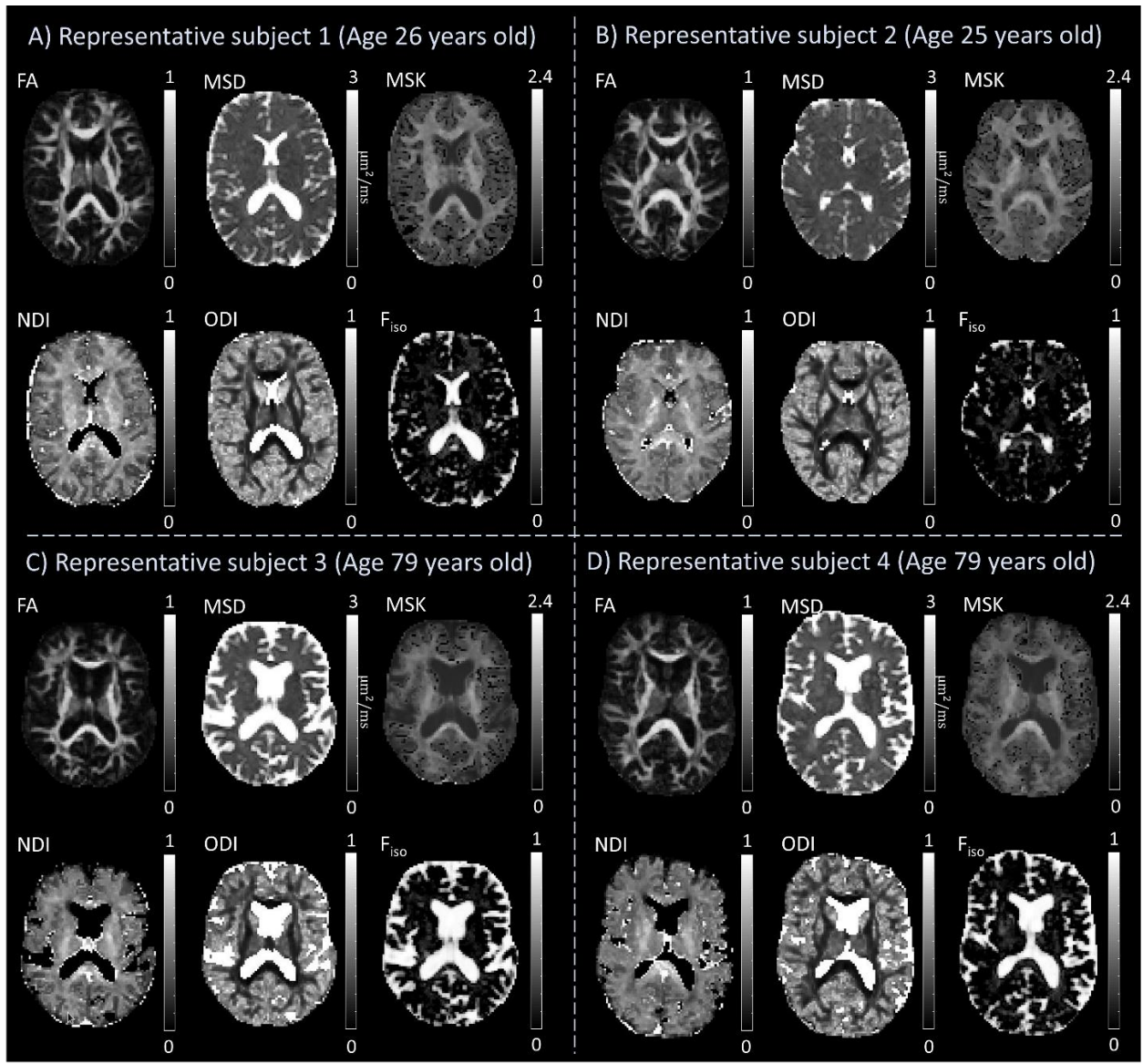

Fig. 1 - Representative maps of the six diffusion MRI metrics (FA, MSD, MSK, NDI, ODI, $F_{iso}$) for two young adults (26 and 25 years old, panels A and B) and for two elders (79 years old, panels C and D). Implausible NODDI estimates in regions containing brain ventricles are removed by setting NDI to 0 and ODI to 1 for voxels with $F_{iso} > 0.9$ (note that estimating NDI and ODI is degenerate for $F_{iso} \approx 1$, c.f. Eq. 2).

## 3.2 Global White Matter dMRI Age Profiles

The mean diffusion metrics computed as the average across the voxels of all WM ROIs are plotted for each metric as a function of age in Fig. 2. FA estimates show a linear decline, with age accounting for approximately 20% of its variance. MSD shows a positively-accelerated effect of age (with 41% of its variance explained by a linear effect, and a further 13% by a quadratic effect), with large increases after 60 years of age. MSK shows a negatively-accelerated effect of age, with a linear effect explaining 25% of its variance, and a quadratic effect explaining an additional 6%, with large decreases after 60 years of age. For the NODDI metrics, NDI also shows a large negatively-accelerated effect, with a linear effect explaining 14% of its variance and a quadratic effect explaining an additional 7%. In contrast, ODI shows only a modest age effect, with a linear effect accounting for 1% of its variance and a quadratic effect accounting for an additional 5%. $F_{iso}$ shows an accelerated increase with age, with linear and quadratic terms explaining 27% and 4% of its variance, respectively. Most metrics show a small effect of sex (approximately 1% of variance) and negligible evidence that the effects of age depended on sex (with linear and quadratic interactions explaining <1%). We therefore drop the sex variable in subsequent analyses.

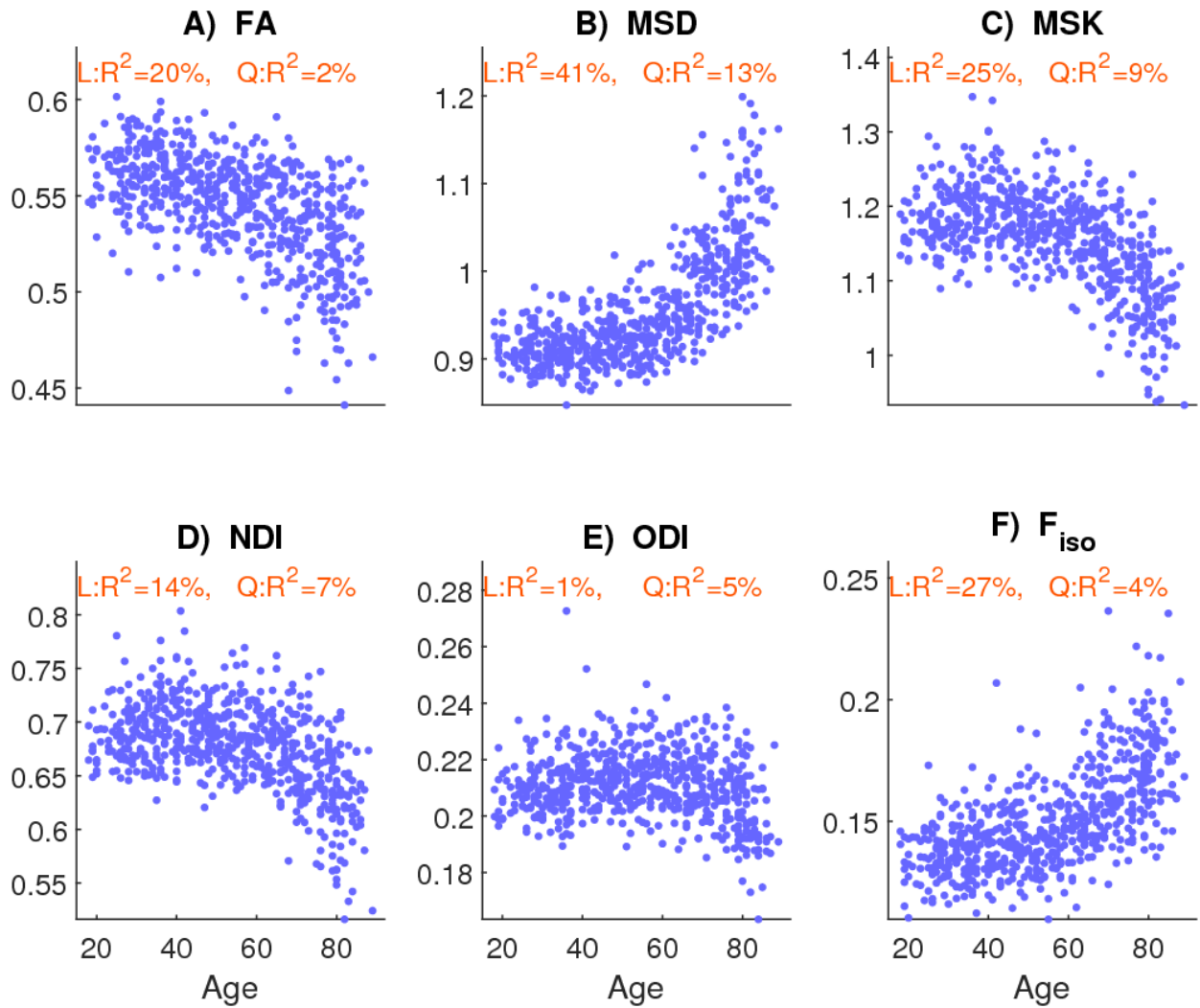

Fig. 2 - Mean diffusion metrics extracted from the voxels of all white matter ROIs as a function of participant's age, for each dMRI metric: A) Fractional Anisotropy (FA) from DTI; B) Mean Signal Diffusion (MSD, in $\mu m^2/ms$) from DKI; C) Mean Signal Kurtosis (MSK) from DKI; D) Neurite Density Index (NDI) from NODDI; E) Orientation Dispersion Index (ODI) from NODDI; and F) Volume Fraction of Free isotropic water diffusion ($F_{iso}$) from NODDI. The proportion of variance ($R^2$) explained by Linear (L) and Quadratic (Q) components of a second-order polynomial fit of age (with covariates of sex and age-by-sex interactions; see text) are shown at the top of each panel.

## 3.3 Regional White Matter dMRI Age Profiles

When splitting the dMRI metrics according to ROI, the resulting $R^2$ values for the second-order polynomial effect of age are shown in Fig. 3. The ROIs are ordered (top-down) according to their mean $R^2$ values across metrics. As expected from the global effects in Fig. 2, the MSD, MSK and $F_{iso}$ metrics tend to show stronger age effects across ROIs than the other three metrics, though there are exceptions: for example, the superior cerebellar peduncle shows stronger effects of age on FA and ODI. Age exerted the biggest effect on: a) anterior brain ROIs such as the Anterior Corona Radiata, Fornix (Column + Body), Corpus Callosum Genu, and the Anterior portion of the Internal Capsule; and b) superior brain ROIs such as the Superior Fronto-Occipital Fasciculus and Superior Corona Radiata. The smallest age effects are observed on the Cerebellar and Cerebral Peduncles, Corticospinal Tracts, Pontine Crossing Tracts and Medial Lemniscus.

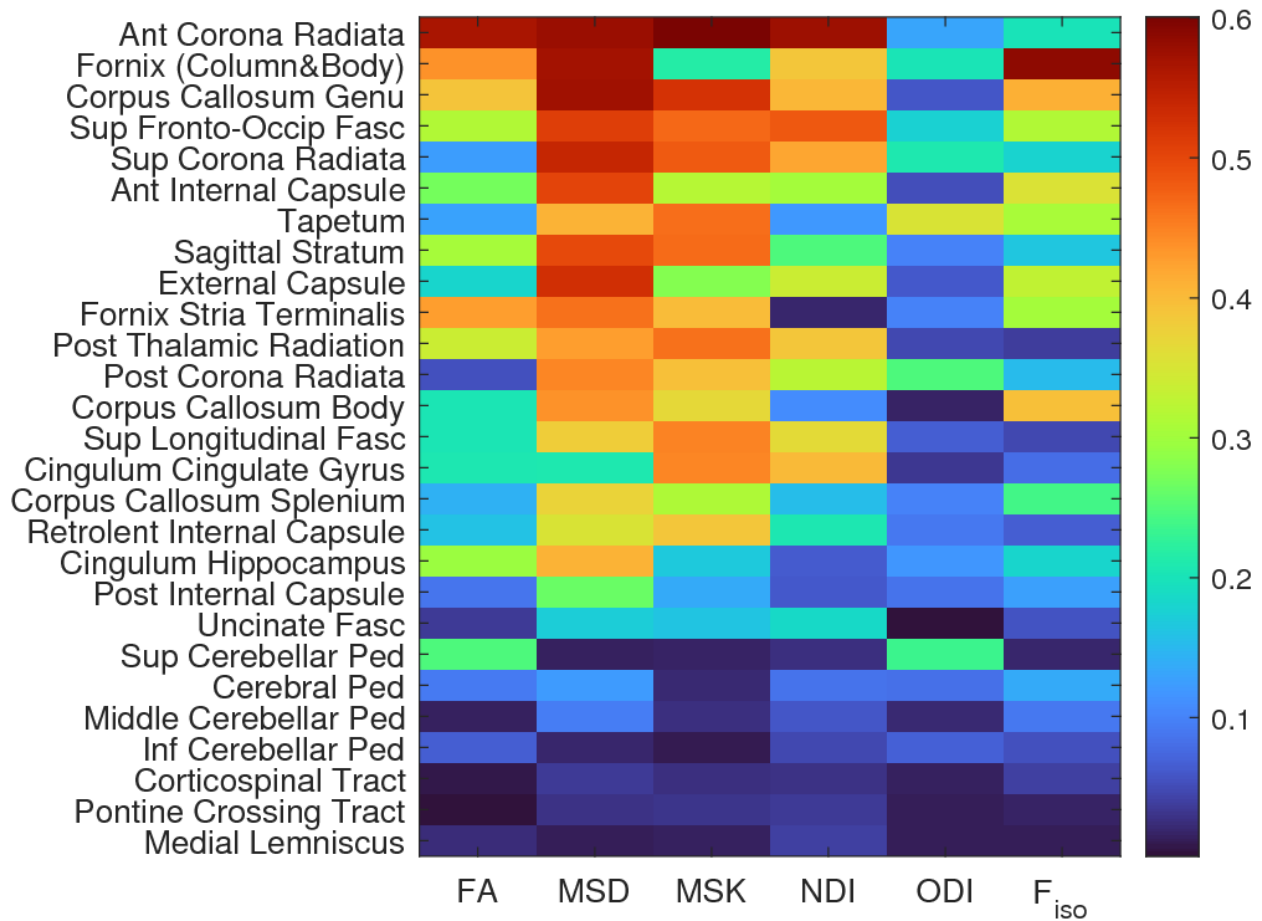

Fig. 3 – Proportion of variance explained ($R^2$) by linear and quadratic effects of age on the six diffusion metrics (FA, MSD, MSK, NDI, ODI, and $F_{iso}$ from left to right) for each ROI separately in different rows. The ROIs are sorted in a descending manner according to their mean $R^2$ values across ROIs. Abbreviations: Ant – anterior; Fasc – Fasciculus; Inf – Inferior, Ped – Peduncle; Post – Posterior; Sup – Superior; Occip - Occipital

## 3.4 Age Effects Within Different Age Subgroups

To further explore the different, nonlinear patterns of age effects across ROIs, Fig. 4 shows linear effects of age for each of the six diffusion metrics across three different age subgroups: 1) subjects aged between 28 and 47 years; 2) subjects aged between 48 and 67 years; and 3) subject aged between 68 and 87 years. ROIs with significant (FDR-corrected) negative and positive effects of age are color-coded by blue and red intensities respectively, while ROIs with non-significant effects are shown in green.

Different age effects across ROIs are apparent from differences across age sub-groups in the linear effect of age within each sub-group. For instance, significant FA declines of the Internal Capsule Posterior limb are only observed for the youngest group (Fig. 4A1), significant FA declines of the External Capsule are only observed for the middle-aged group (Fig. 4A2), while significant FA declines in the Corpus Callosum Body and Splenium are only observed in the oldest sub-group (Fig. 4A3).

MSD shows the strongest effects of age in the oldest sub-group across a large number of ROIs (Fig. 4B3). For MSK, positive age effects are observed for the External Capsule and the Posterior limb of the Internal Capsule for the youngest sub-group (Fig. 4C1), while declines in the middle-aged sub-group are only observed for some WM ROIs (Fig. 4C2), particularly for the Corpus Callosum Genu, the Anterior Corona Radiata, the Retrolenticular portion of the Internal Capsule, the Posterior Thalamic Radiation, and the Superior Longitudinal Fasciculus. For the oldest age-group, almost all WM ROIs show negative MSK variation rates (Fig. 4C3).

NDI shows similar trends to MSK in most of the WM ROIs. For example, like MSK, positive and negative NDI rates are observed in the Internal Capsule Posterior limb and Posterior Thalamic Radiation for the younger age-group (Fig. 4D1), while significant NDI decreases are observed for almost all the ROIs that showed significant MSK negative rates (Fig. 4D1 and 4D2).

The clearest example of a quadratic relationship with age is for ODI, where the youngest sub-group shows significant positive age effects for many ROIs (e.g., Corpus Callosum Genu, Internal Capsule, Anterior Corona Radiata and Longitudinal Superior Fasciculus; Fig. 4E1), whereas the middle-aged sub-group shows few effects of age across ROIs (Fig. 4E2), and the oldest sub-group shows significant *negative* effects in most ROIs (Fig. 4E3).

Different age-subgroups show significant positive $F_{iso}$ variation rates for different WM regions (Fig. 4F), whereas negative $F_{iso}$ variations are only observed in the Cingulum Cingulate Gyrus for the oldest sub-group (Fig. 4F3).

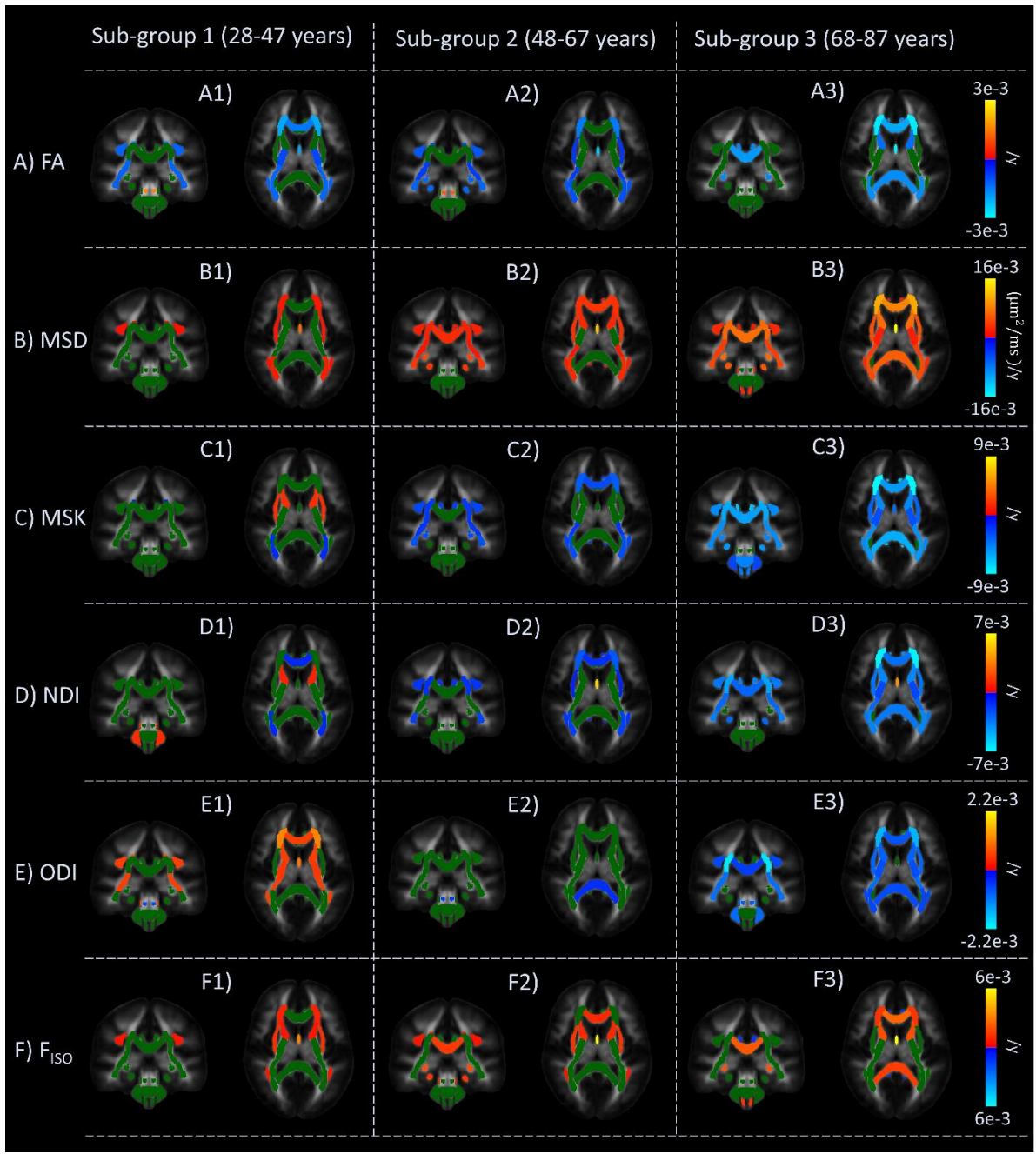

Fig. 4 – Linear effects of age for the six diffusion metrics (FA, MSD, MSK, NDI, ODI, and $F_{iso}$ from panels A to F) within three age sub-groups (left to right subpanels), overlaid on the JHU-ICBM FA template - in each panel results are displayed for a coronal (right) and an axial (left) slice. Correction for multiple-comparison is performed using FDR (q<0.05). Significant negative and positive age effects are color-coded by blue and red intensities respectively; while non-significant effects are shown in green.

## 3.5 Correlations Between dMRI Metrics and Factor Analysis

Fig. 5 shows the Pearson correlation coefficient between each pair of the six metrics. The upper triangle shows the raw correlations, while the lower triangle shows the partial correlation having removed linear and quadratic effects of age. The similarity of the two triangles indicates that the correlations between metrics are not driven primarily by common age effects (i.e, reflect differences between individuals and regions beyond those due to age). Higher positive correlations are observed between MSK and NDI and between MSD and $F_{iso}$, while the strongest negative correlation is between FA and ODI. Correlations near zero are observed between MSD and ODI, between MSK and $F_{iso}$, and between MSK and ODI when linear and quadratic effects of age are removed.

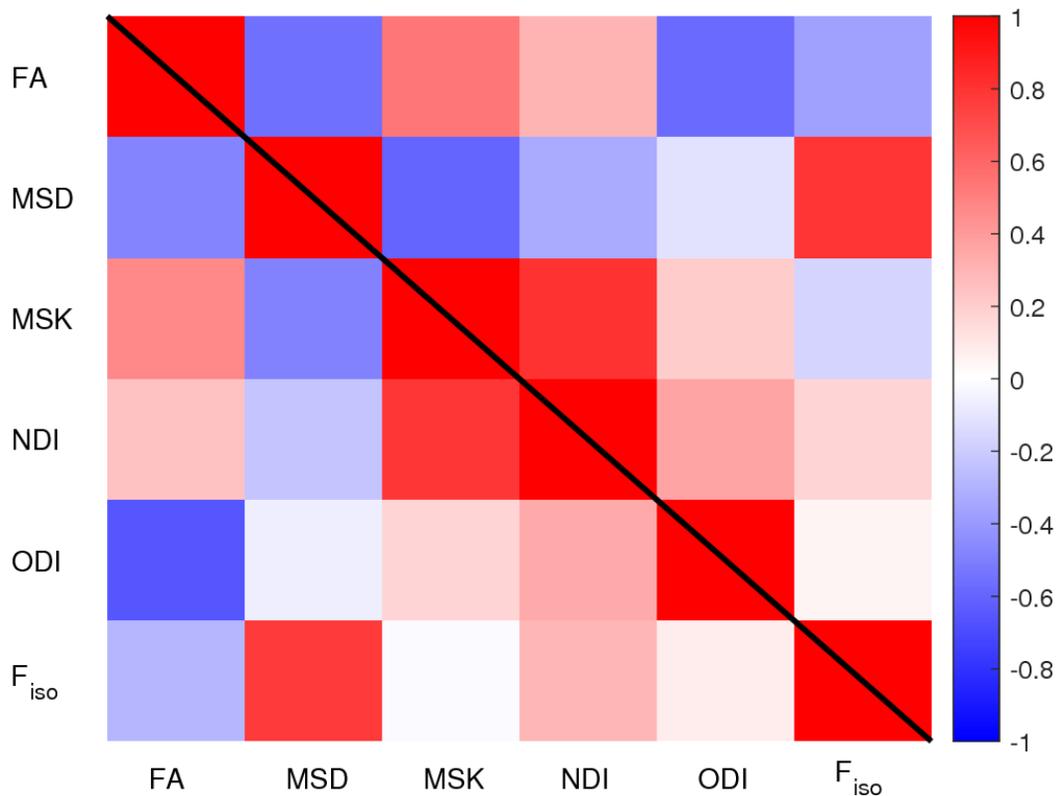

Fig. 5 – Pearson correlation coefficient (R) between each pair of the metrics. The upper right triangle shows raw correlations; the lower left triangle shows correlations after removing linear and quadratic effects of age.

The first three principal components (PCs) of the above correlation matrix explain 46.3%, 29.0% and 20.3% of the variance respectively, with the fourth PC only explaining 2.2%. Therefore, only the first three dimensions are retained, but rotated to maximise variance of the loadings (i.e, factor analysis). The factor scores across metrics are shown in the three upper panels of Fig. 6. The first factor loads most strongly and positively on MSK and NDI, with a smaller negative loading on MSD and smaller positive loadings on the rest, particularly FA. Thus, this first factor most likely reflects non-Gaussian diffusion effects in WM tissues (i.e., the degree to which the paths of water molecules are hindered/restricted by biological obstacles of WM). The second factor positively loads on MSD and $F_{iso}$, most likely reflecting the free water contribution to the diffusion-weighted signal. The third factor only has strong positive and negative loadings on ODI and FA respectively, likely reflecting effects of tissue configuration complexity such as presence of crossing, dispersing or fanning fibres. The lower three panels of Fig. 6 show how the factor loadings across participants (averaged across ROI) vary with age: Factor 1 shows an inverted U-shape with age (linear and quadratic effects explaining 29% and 17% of its variance), Factor 2 shows a positively-accelerated effect of age (with 54% of its variance explained by a linear effect, and a further 16% by a quadratic effect), while Factor 3 shows a linear increase with age, explaining 11% of its variance. Given these interpretations of the three factors, Fig. 6 reinforces how FA and MSD are likely to be influenced by a mixture of underlying factors, whereas the three NODDI metrics are largely selective to each factor, and MSK loads predominantly on Factor 1.

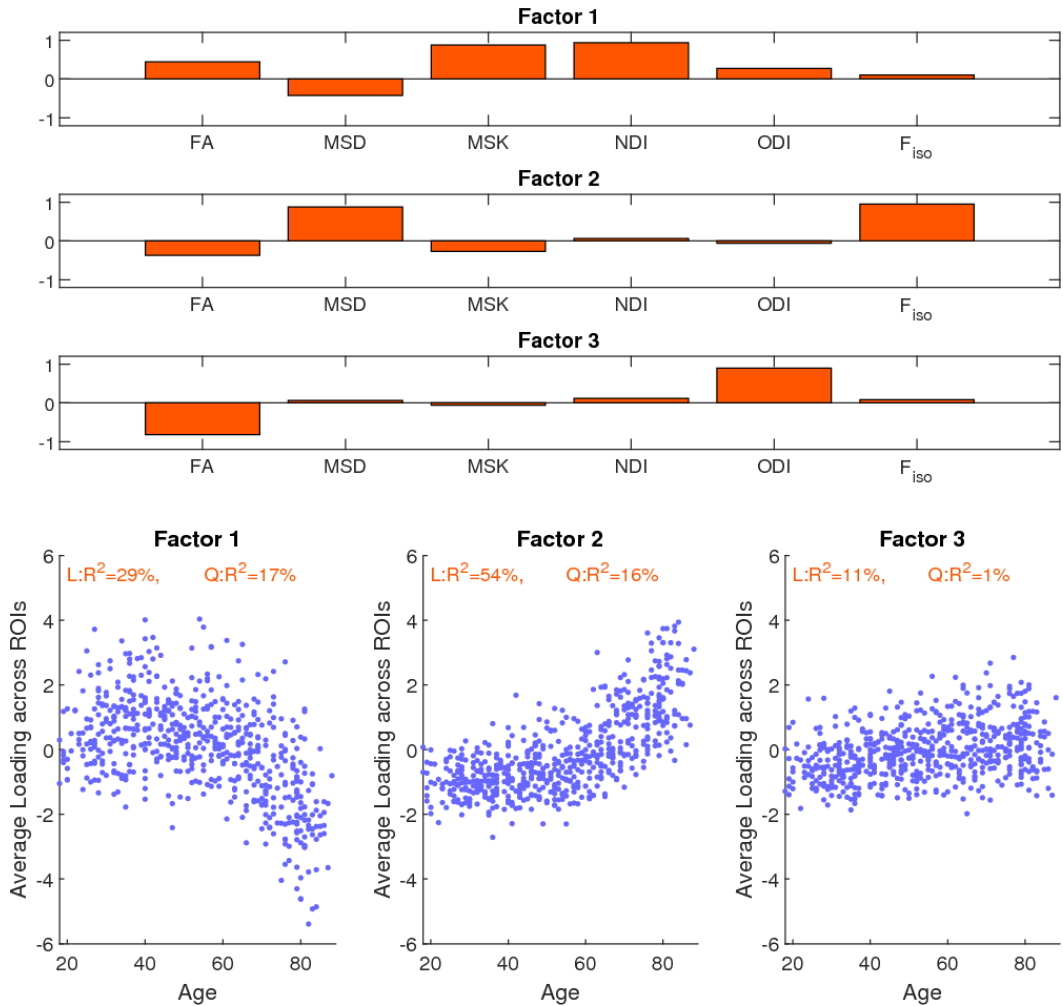

Fig. 6- Loadings of three factors from factor analysis across the six diffusion metrics (upper three panels) and their profiles against age (lower three panels).

The loadings of each factor across ROIs are shown in Fig. 7, which seem to be aligned with the expected microstructural features of different WM regions. For example, the three ROIs in the Corpus Callosum, where the underlying fibre architecture is characterised by a single fibre population, have relatively low loadings on Factor 3, while the ROIs in Corona Radiata have high loadings on this factor, consistent with increased fibre complexity in that region, including crossing fibres. For Factor 2, ROIs close to cerebrospinal fluid such as the Column and Body of the Fornix, Corticospinal Tract, Medial Lemmiscus, Cerebellar Peduncles, Cerebral Peduncle have relatively high loadings compared to regions further away from the ventricles, such as the Cingulum Cingulate Gyrus, Cingulum Hippocampus and the Fornix Stria Terminalis.

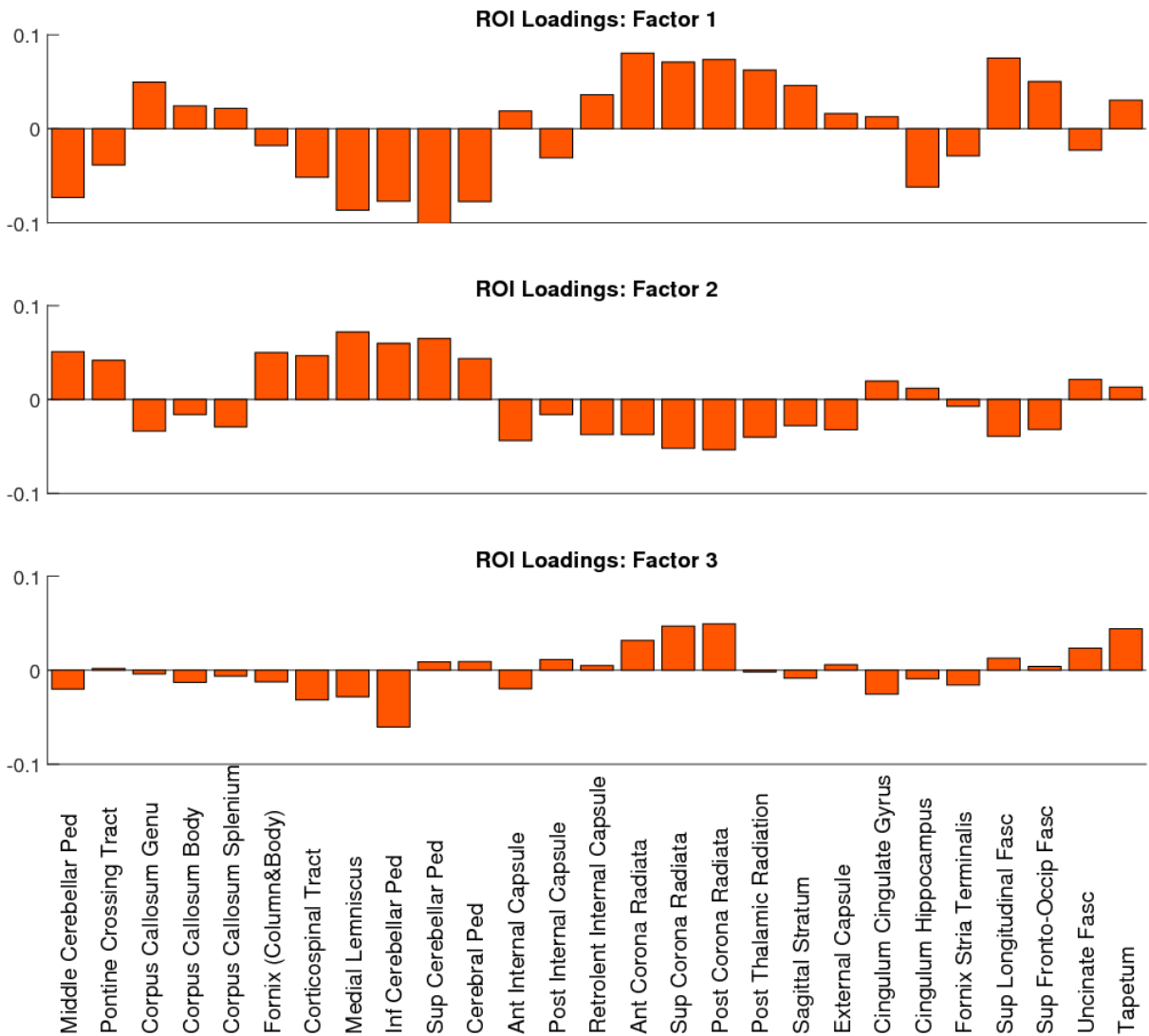

Fig. 7 - Loadings of three factors from factor analysis on each of the 27 ROIs.

Finally, the effect of age on each Factor in each ROI is shown in Fig. 8. The strongest effects with age are observed for Factors 1 and 2, which show effects of age on most ROIs, while for Factor 3 strong effects with Age are only observed for a handful of ROIs. As expected from Fig. 3, a few ROIs, including the Cerebellar Peduncles, the Pontine Crossing Track, Corticospinal Tract and Medial Lemniscus, show weak effects of age on every factor.

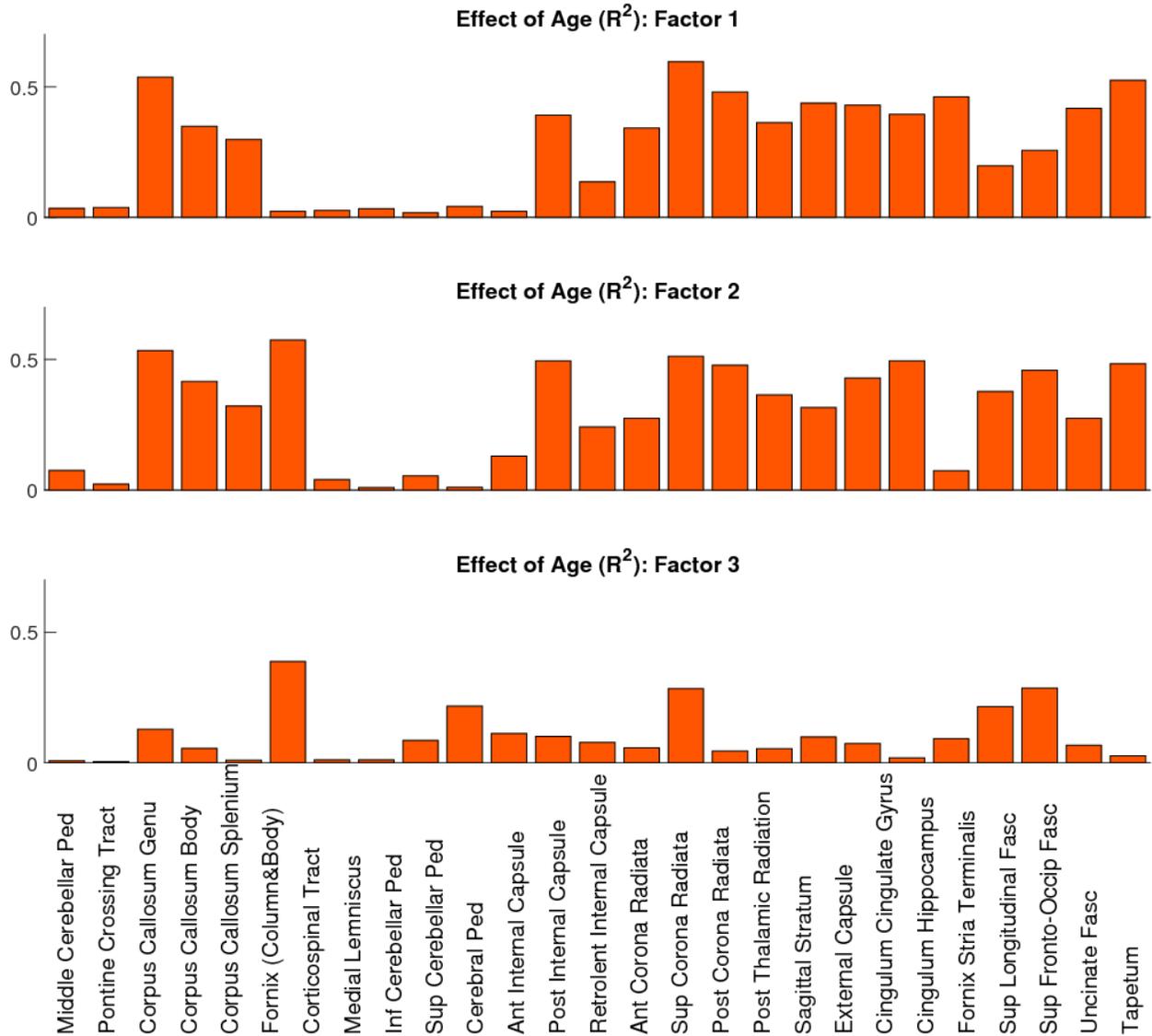

Fig. 8 - Effects of Age ($R^2$ from second-order polynomial fit) for each factor and each of the 27 ROIs.

# 4. Discussion

Previous studies showed that dMRI can reveal information about age-related microstructural alterations of brain tissues that are not detected by conventional imaging techniques (Maillard et al., 2013; Nusbaum et al., 2001; Pelletier et al., 2017). While conventional structural MRI contrasts show that in general the volume of WM decreases only after the fifth decade of life (Bethlehem et al., 2022; Lebel et al., 2012; Walhovd et al., 2011), Diffusion Tensor Imaging (DTI) suggests that diffusion Fractional Anisotropy (FA) in WM regions declines with age from early adulthood (Davis et al., 2009; Lebel et al., 2012; Pfefferbaum et al., 2000; Sullivan et al., 2001). These studies hypothesised that FA declines were associated with degenerative processes such as axonal loss and demyelination; however such interpretation is limited by DTI's lack of specificity (De Santis et al., 2014; Wheeler-Kingshott et al., 2009). In recent years, more advanced dMRI techniques have been applied in an attempt to provide more specific information on white matter microstructural changes (Beck et al., 2021; Billiet et al., 2015; Chang et al., 2015; Coutu et al., 2014; Cox et al., 2016; Das et al., 2017; Falangola et al., 2008; Kodiweera et al., 2016; Merluzzi et al., 2016). However, results across different studies do not always agree, which could be a consequence of 1) different dMRI techniques used and 2) different demographic characteristics of the populations studied.

In this study, we addressed these discrepancies by comparing different dMRI techniques (DTI, DKI, NODDI) on a cohort of adults approximately uniformly distributed across the ages 18-88. Firstly, we found significant quadratic effects for most of the diffusion MRI metrics considered (especially for MSD, MSK, NDI and $F_{iso}$) in both whole-brain and regional white matter (Fig. 2 and Fig. 3), consistent with previous reports (Beck et al., 2021; Billiet et al., 2015; Coutu et al., 2014; Cox et al., 2016; Falangola et al., 2008; Lebel et al., 2012; Yeatman et al., 2014). We further explored these quadratic effects by looking at different brain regions over three age sub-groups (Fig. 4). The different (linear) effects of age in each sub-group imply that the age-related patterns observed across metrics and ROIs is highly dependent on the age ranges of the volunteers included in a study, which might explain some of the inconsistencies in previous studies using DKI/NODDI (Billiet et al., 2015; Chang et al., 2015; Cox et al., 2016; Merluzzi et al., 2016). Overall, our results also confirm that going beyond DTI, more advanced dMRI techniques based on signal representation (e.g., DKI) and microstructural models (e.g., NODDI) can provide different information about microstructural age-associated changes (Fig. 4 and Fig. 5), consistent with prior claims (Billiet et al., 2015; Chang et al., 2015; Coutu et al., 2014; Das et al., 2017; Falangola et al., 2008; Gong et al., 2014; Kodiweera et al., 2016; Lätt et al., 2013). However, our factor analysis shows that variation in all six diffusion metrics used in this study can be captured by just three main dimensions (Fig. 6), which we have linked to effects of 1) non-Gaussian diffusion, 2) tissue configuration complexity and 3) free-water content. The loading of these factor across different WM ROIs was shown to be aligned with their expected microstructural differences (Fig. 7) and reveals regional differences in age-related changes (Fig. 8). These aspects are discussed in more detail bellow.

### 4.1 FA has Limited Specificity to Age-related Changes

Consistent with early DTI ageing studies, e.g. (Davis et al., 2009; Lebel et al., 2012; Pfefferbaum et al., 2000; Sullivan et al., 2001; Zhang et al., 2010), the results of the present study show general WM FA declines from age 18 onwards (Fig. 2). However, these early effects of age are not found for the diffusion metrics that are invariant to fibre architecture (i.e. MSD, MSK, NDI). Therefore, as mentioned in previous studies (Billiet et al., 2015; Chang et al., 2015; Kodiweera et al., 2016), early age-related changes of FA are likely to be a consequence of changes in fibre architecture that can be detected by NODDI's orientation dispersion. Although ODI showed only a modest age effect in the global WM profiles (Fig. 2), the impact of ODI in FA is highlighted by the strong negative correlations between these metrics in Fig. 5. Moreover, the poor specificity of FA is supported by our factor analysis (Fig. 6), which suggests that FA estimates reflect a mixture of all three factors detected in this study, with Factor 3, the one related to fibre architecture alterations, showing the highest loadings.

The poor specificity of FA limits its use in the interpretation of age-related microstructural changes, as exemplified in our age subgroup analysis. For instance, while early studies interpreted FA declines as WM degeneration, results from the youngest age subgroup show that declines in FA (Fig. 4A1) are not accompanied by declines in MSK and NDI (Fig. 4C1 and Fig. 4D1). Instead, these early FA decreases are in the line with significant ODI increases observed in some WM regions, such as the Internal Capsule Posterior Limb, Anterior Corona Radiata and Corpus Callosum Genu (Fig. 4E1). In addition to its poor specificity to detect late maturation processes, FA is inadequate in predicting WM degeneration in older age. Indeed, while both MSK and NDI show widespread declines in older age (Fig. 4C3 and Fig. 4D3), FA declines are only observed in some WM ROIs (Fig. 4A3). This is likely a consequence of the widespread decrease in tissue configuration complexity, as measured by ODI decreases (Fig. 4E3), which has the opposite impact of true WM degeneration in FA estimates.

Given FA's lack of specificity, our results suggest that FA is not an adequate WM marker for use in future studies assessing, for example, the relationships between brain properties and age-related cognitive declines. This observation also provides strong evidence to support previous claims about the limited sensitivity and specificity of DTI (De Santis et al., 2014; Henriques et al., 2021a, 2015; Jeurissen et al., 2013; Wheeler-Kingshott et al., 2009).

### 4.2 Decoupling Age-Changes from Fibre Dispersion Confounds

One main achievement of more advanced dMRI techniques that go beyond DTI is the ability to decouple microstructural alterations from confounding effects related to fibre architecture. In ageing studies, minimising this confounding effect is important since alterations of the morphology or dispersion of white matter bundles are likely to be directly related to the expected macroscopic volume changes of WM observed across the human lifespan (Bethlehem et al., 2022; Lebel et al., 2012; Walhovd et al., 2011). Indeed, DKI metrics independent of fibre architecture can be obtained from diffusion-weighted signals averaged along multiple directions (also known as powder-averaged signals) (Henriques et al., 2021a, 2019; Jensen et al., 2005). On the other hand, microstructural models have been designed to separate effects of microstructural features from fibre orientation distribution properties (Jespersen et al., 2010, 2007; Kaden et al., 2016a; Novikov et al., 2018), as in NODDI (Zhang et al., 2012).

The results obtained from the present cohort show that metrics designed to be independent from fibre architecture (e.g. MSK and NDI) only present declines from the late 40s (Fig. 2 and Fig. 4). This suggests that these metrics are sensitive to late maturation processes not resolved by DTI, and their later declines may be more specific to general degeneration processes than DTI metrics such as FA. It should be noted that, although MSD is also expected to be invariant to fibre dispersion effects (c.f. Fig. 5), this metric provides less specific tissue microstructural characterization since it is highly affected by increases in free water partial volume effects with age (c.f. Fig. 5 and Fig. 6). This may also explain the increases of MSD observed from the two younger age sub-groups (c.f. Fig. 4B1 vs Fig. 4F1 and Fig. 4B2 vs Fig. 4F2).

## 4.3 MSK and NDI sensitivity to general maturation and degeneration processes

Even if techniques can separate microstructural alterations from confounding effects of changes in fibre dispersion, DKI and NODDI may not be adequate to resolve other microstructural changes, such as distinguishing axonal loss from demyelination. For instance, MSK quantifies alterations in the degree of non-Gaussian diffusion in tissues. Our results indicate that MSK has a low loading on Factors 2 and 3 (Fig. 6), indicating that this measure is not strongly associated to fibre dispersion effects nor gross increases of free water contamination effects; however, since remaining tissue non-Gaussian alterations can be a consequence of different underlying WM maturation/degeneration mechanisms, MSK changes cannot be used to discriminate specific microstructural alterations.

The same limitation applies to NODDI: while ODI and $F_{iso}$ may be mainly associated with gross fibre dispersion and free water contamination effects respectively, this study does not provide evidence that NDI can accurately target changes in axonal density. Although NDI from NODDI was designed to estimate neurite density, its interpretation may be limited by over-simplistic biophysical assumptions, as pointed by previous studies (Henriques et al., 2019; Lampinen et al., 2017; Novikov et al., 2018). Indeed, the high correlations between NDI and MSK observed in our present analysis (Fig. 5) suggest that, like MSK, NDI cannot distinguish between specific degeneration mechanisms like axonal loss versus demyelination. This requires more advanced diffusion MRI techniques (c.f. Section "4.6 Limitations and Future Directions").

Despite this general pitfall, due to their general sensitivity to maturation and degeneration process decoupled from fibre dispersion and free water contamination effects, both MSK and NDI may provide better quantities than DTI metrics to study the relationship between brain tissue microstructure and cognitive differences (Arfanakis et al., 2016; Han et al., 2016; Huber et al., 2019), when more acute process of tissue damage can be ignored, such as cytotoxic and vasogenic oedema effect that occurs in ischemic stroke lesions or traumatic brain injury, e.g. (Alves et al., 2022; Hui et al., 2012; Kamiya et al., 2020a; Rudrapatna et al., 2014; Zhuo et al., 2012).

## 4.4 Comparison with Previous dMRI Studies

As mentioned above, previous studies using NODDI describe inconsistent effects of age. The results of this study show that correlations between NODDI metrics and subject age depend on the age ranges (c.f. Fig.

4), suggesting that previous inconsistencies in the literature are a consequence of differences in sample distributions of age. For example, the positive NDI variation rates observed by Chang et al. (2015) and Kodiweera et al. (2016) might be a consequence of a large number of young to middle-aged participants, while the negative NDI variation rates observed by Cox et al. (2016) and Merluzzi el al. (2016) might reflect the relatively older age ranges included in their respective cohorts. The wider and more uniform distribution of age in the present Cam-CAN cohort allows the detection of age periods where NDI both increases and then declines (c.f. Fig. 2 and Fig. 4).

In this study, higher $R^2$ values for the age effects are present for $F_{iso}$, MSD and MSK when compared with FA, NDI and ODI (Fig. 3). Higher $R^2$ values for MSD agree with the higher MD $R^2$ values reported by Cox et al. (2016). However, these changes may be difficult to interpret due to the lack of specificity of MSD/MD. For example, we found strong correlations between MSD and $F_{iso}$ (Fig. 4), and previous studies have shown reduced associations between MD and age when MD is corrected for free-water contamination (e.g. Chat et al. 2018). In a recent study (Pieciak et al., 2022), $F_{iso}$ was shown to explain most of DTI changes with age, which is in line with the higher $R^2$ values shown here (Fig. 3). However, this study did not assess effects from non-Gaussian diffusion, which here is shown to explain 46.3% of the age variance (Fig. 6).

Although in this study we show that MSK and NDI age-profiles are generally consistent (Fig. 2, Fig. 4, Fig. 5, and Fig. 6), the same was not observed in all previous studies (e.g. Billiet et al., 2015). This discrepancy is likely a consequence of different methodologies used. For instance, while the diffusion metrics of our study are extracted from ROIs in each participant's native space, the diffusion metrics extracted by Billert et al. (2015) were obtained after warping and reslicing images to a common template. While the analysis preformed in the present study was designed to minimize the effect of free water partial volume increase with age (i.e. by the exclusion of ROI voxels that mainly contain free water, $F_{iso} >$ 0.9), the interpolation entailed by reslicing may have the opposite effect of highlighting age-related increases on free water fraction estimates. Template registration may also explain the poor sensitivity of DKI to age alterations reported by Billert et al. (2015), since interpolation may induce the propagation of inaccurate kurtosis estimates, given that implausible high magnitude kurtosis estimates have been reported in previous studies (Henriques et al., 2021b; Tabesh et al., 2011). In our study, in addition to avoiding detrimental effects of diffusion metric map registration, the use of powder-averaging was shown to successfully mitigate implausible negative kurtosis in WM brain regions (c.f. Fig.1).

Regarding the study by Beck et al. (2021), which used a cohort of participants with a similar number to our study, their age profiles of different dMRI metrics extracted from global WM skeletons agree with our global WM dMRI metrics age profiles. However, some differences in ODI profiles can be noted. While Beck et al. (2021) showed only slowing down of the rate of ODI increase in older age, our analysis suggests that ODI may actually decrease in older age (Fig. 4E2 and Fig. 4E3). These discrepancies may likely be explained by differences on the WM regions of interest assessed – the thinner WM skeletons used by Beck et al. (2021) are likely to be less sensitive to ODI decreases than the wider WM regions of interest used in our study. Despite this, the work by Beck et al. (2021) agrees that age profiles from different dMRI metrics may be similar. For example, the similarities observed between their NDI and mean kurtosis profiles are in line with the similarities observed between our NDI and MSK estimates. Nonetheless, our study goes further by providing a formal analysis on information redundancy across different dMRI metrics, leading to a more comprehensive understanding and interpretation of their relationships in both global and regional WM regions (see the discussion of our factor analysis below).

Finally, a recent study using advanced diffusion encoding to resolve sources of non-Gaussian diffusion (anisotropic vs isotropic kurtosis) revealed that MSK decreases in old age are in line with anisotropic kurtosis decreases (Kamiya et al., 2020b). These results support the hypothesis that age-related MSK decreases are, most likely, related to general white-matter degeneration, rather that increases of free water partial volume effects (captured by isotropic kurtosis).

## 4.5 Factor Analysis Across Metrics

Our factor analysis across subjects and metrics supports three main dimensions (Fig. 6), explaining a total of 97.8% of the variance in the data. A similar analysis by Chamberland et al. (2019), using a different range of metrics, reported only two principal components (PCs), with interpretations similar to Factors 1 and 3 in the present study. However, the two components reported by Chamberland explained only 80% of the variance in their data, with no other PCs reported. It is therefore unknown whether a 3$^{rd}$ component corresponding to metrics of free-water was also present in that study, since the authors did not include metrics from dMRI techniques designed to decouple such affects. Moreover, free water contributions may be expected to explain less variance in the study by Chamberland and colleagues than in the present study given their much smaller range of ages.

Factor 1 shows an inverted U-shaped profile with age, explaining 46.3% of the variance, suggesting it is sensitive to ongoing maturation processes into the early 30s, as well as white-matter degeneration later in life. As discussed above, this is likely to reflect general mechanisms of tissue maturation/degeneration related to myelination and axonal density. Factor 2 shows a positive correlation with age, which accelerates from the 60s and explains 29.04% of the variance. This is consistent with enlarged interstitial spaces with age, which results in increased partial volume effects from CSF, as captured by $F_{iso}$ and MSD (which here is not corrected by free water effects in DKI modelling). Finally, Factor 3 shows a linear increase with age, which explained 20.3% of the variance. This is consistent with a decrease in orientational coherence of the underlying white matter fibres with age as captured by ODI and FA. The strong positive correlation between $F_{iso}$ and MSD (Fig. 4) suggests that when free-water modelling is not performed (standard DTI fitting), the impact of this increased partial volume effect is captured by an increase in MSD.

Different ROIs show different effects of age, and different loadings on the three factors. For example, Factors 1 and 2 show the well documented 'anterior-posterior gradient of ageing' across the three ROIs covering the corpus callosum (genu, body, and splenium), with a stronger association with age in the genu (Fig. 8). Factor 3 shows in general weaker correlations with age, with only three ROIs (Fornix Column and Body, Superior Corona radiata, and Superior Fronto-Occipital Fasciculus) showing an $R^2$ value greater than 0.25. This suggests that factors that affect tissue non-Gaussian diffusion properties (such as fibre density or myelin) and free water partial volume effects change more with age than fibre orientation complexity.

## 4.6 Limitations and Future Directions

In this study, we did not consider directional metrics of DTI and DKI such as the axial and radial diffusivities, and the axial and radial kurtosis, since their interpretation may be limited to WM regions comprising of single fibre populations (De Santis et al., 2014; Henriques et al., 2015; Jeurissen et al., 2013; Wheeler-Kingshott et al., 2009). Although our study mainly focused on advanced diffusion MRI metrics that can be generally applied to different WM fibre configurations, an intrinsic pitfall of this work is that it does not explore if directional DTI/DKI metrics can provide additional information that could potentially be used, for example, to distinguish mechanism of fibre loss and demyelination as suggested by previous studies (Beck et al., 2021; Coutu et al., 2014; Das et al., 2017; Davis et al., 2009; Fieremans et al., 2013; Gong et al., 2014; Lätt et al., 2013). Therefore, in future studies, it will be of interest to perform similar factor analysis on directional metrics of DTI/DKI or other metrics from models that use information from directional quantities, such as the White Matter Tract Integrity model (a two-compartmental microstructural model that can be reconstructed from the full DKI tensor under the assumption that fibres are well-aligned; Fieremans et al., 2013, 2011; Henriques et al., 2021a).

Although the present study focused on the most commonly used dMRI techniques in previous ageing studies (e.g. DTI/DKI/NODDI), future studies could check if metrics from other microstructural models that estimate a larger number of parameters can also be reduced to the three main factors detected in this study, including metrics from the Composite Hindered and Restricted Model of Diffusion (CHARMED; Assaf and Basser, 2005), Neurite Orientation Dispersion and Density Imaging with Diffusivities Assessment (NODDIDA; Jelescu et al., 2016), and the general standard model (SM) for WM (Novikov et al., 2018). These models were not considered here since they are known to be ill-posed on current conventional diffusion MRI acquisitions, consequently being more prone to inaccurate and imprecise estimates (Jelescu et al., 2020, 2016; Novikov et al., 2018) and requiring more complex fitting routines (Coelho et al., 2022; Mozumder et al., 2019; Reisert et al., 2017). Other models that are well-posed include the one- and two-compartment spherical mean techniques (Kaden et al., 2016b, 2016a), but these were not considered here because they were already shown to provide the same information as the DKI quantities explored here (Henriques et al., 2019). Future studies could expand our analyses to diffusion MRI techniques that use additional MRI information from diffusion-weighted data with higher b-values and different diffusion timing parameters (e.g. diffusion pulse separation Δ and diffusion pulse duration δ; Jensen et al., 2016; Jespersen et al., 2010, 2007; Palombo et al., 2020; Veraart et al., 2020), advanced diffusion encodings, (Alves et al., 2022; Eriksson et al., 2013; Henriques et al., 2021c, 2020; Kerkelä et al., 2020; Lasič et al., 2014; Novello et al., 2022; Shemesh et al., 2012, 2011; Shemesh and Cohen, 2011; Szczepankiewicz et al., 2019, 2016, 2015; Topgaard, 2017), and/or different relaxation times (Anania et al., 2022; Slator et al., 2021; Veraart et al., 2018).

Regarding the characterization of the age-profiles for different metrics, in the present study these are characterized using quadratic and linear regression models. Although these polynomial models can detect the presence of global age-related changes, more sophisticated methods like splines may provide more accurate estimates of age-related trajectories, particularly if tissue maturation occurs at faster rates than the rates of tissue degeneration (Fjell et al., 2010; Lebel et al., 2012; Yeatman et al., 2014). Thus, while the present focus was on comparing the effects of age on different dMRI metrics, rather than making strong claims about the neuroscience of ageing, future studies could employ more sophisticated methods for estimating age trajectories, particularly inflection points when rates of WM change from increasing to decreasing. Furthermore, our results are all derived from cross-sectional differences in age across individuals; future studies need to compare them with results from longitudinal dMRI datasets (e.g., Barrick et al., 2010; Beck et al., 2021; Sexton et al., 2014; Vik et al., 2015), where age can be properly dissociated from year of birth. Finally, although here we focus on dMRI metrics in WM regions of interest, future studies could extend our analyses to the characterization of age differences in grey-matter (Falangola et al., 2008; Gong et al., 2014; Helpern et al., 2011b).

# 5. Conclusion

This study provides a better understanding of the relationship between different dMRI models and their sensitivity to age-related changes. While we confirm that the sensitivity and specificity of fractional anisotropy from "standard" DTI is limited by white matter fibre dispersion/crossing confounding effects, we show that advanced models reveal additional insights, since these are capable of separating age-related microstructural information from mesoscopic tissue alterations (e.g. changes on the fibre dispersion or crossing degree). Factor analysis across six diffusion metrics revealed only three factors, which are likely to reflect: 1) white matter maturation followed by degeneration processes; 2) increase in free water partial volume effects with accelerated increases from the 60s; 3) more subtle alterations in fibre organization (i.e. changes in fibre crossing and dispersion). While FA was shown to reflect a combination of all three factors, both MSK and NDI aligned with factor 1, while $F_{iso}$ and ODI aligned with Factors 2 and 3 respectively. Finally, the three different factors show different loadings in different white matter regions, revealing that age alterations have regional effects that reflect distinct combinations of different underlying microstructural alterations.

## Disclosure statement

The authors have no actual or potential conflicts of interest.

# Supplementary Material

# A. Quantification of images corrupted by motion-related stripe artefacts

The two phases used to quantify diffusion-weighted volumes corrupted by motion artefacts are described in this section.

  Phase 1: Head motion during data acquisition not only induces volume misalignments but also signal loss due to abrupt tissue displacement (e.g., Tournier et al., 2011). For axial acquisitions, as is the case for the data acquired on the Cam-CAN project, this can be qualitatively observed along sagittal slices in the form of single slice signal loss or in the form of a "striping" pattern of intensities (Tournier et al., 2011). For the present study, this type of artefact is quantified by convolution of each sagittal slice of all diffusion-weighted volumes with the kernel shown in step 1, Fig. A.1. Fig. A.1 shows the effect of the kernel convolution on images that are (panel A) and are not (panel B) corrupted by motion artefacts. As the figure shows, resulting images present greater intensities if raw images were corrupted by this type of artefact (step 2 in Fig. A.1). Voxel intensities were then added to generate a single value for each volume (step 3 in Fig. A.1). This measure is designated here as the apparent "stripe" index ($aSI$), since its magnitude will not only depend on the amount of motion artefacts, but also on morphological characteristics of the brain of each subject (e.g. larger brains have more brain voxels) and on the diffusion-weighted parameters (e.g. higher tissue intensities on lower b-values are intrinsically associated with higher contrasts between tissue intensity and signal loss and thus larger values of $aSI$). To remove the latter two dependencies, $aSI$ values for the same subject and b-value are grouped and normalized by its group minimum (step 4 in Fig. A.1), resulting in values equal or larger than 1, which we defined as the corrected stripe index $cSI$.

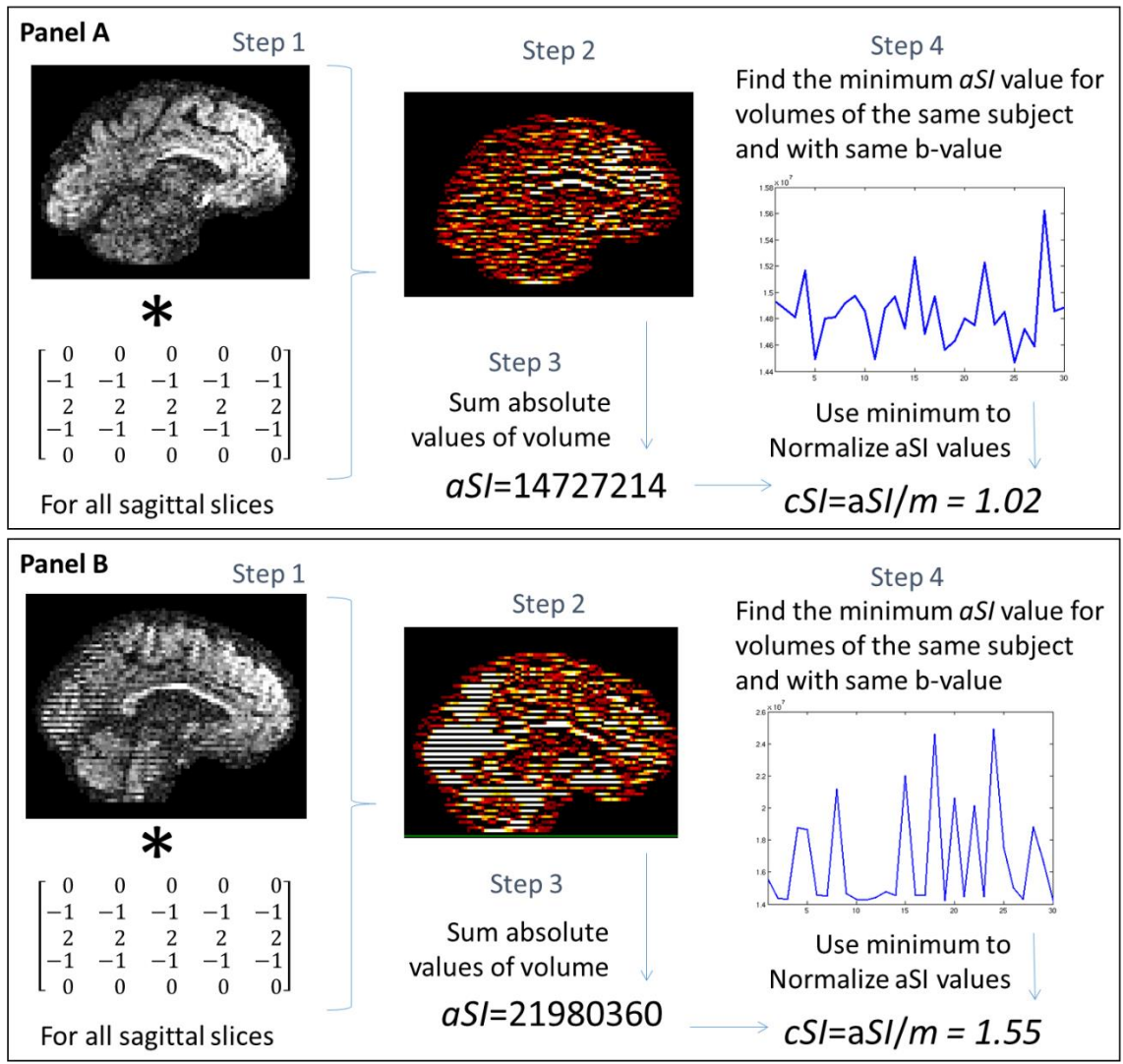

Fig. A.1 – First phase of the procedure to quantify the artefacts induced by abrupt subject motion during data acquisition. On panel A the procedure is illustrated for a volume not corrupted by motion artefacts, while panel B for a volume that visually revealed evident stripe artefacts.

Phase 2: From empirical observations, volumes with cSI larger than 1.25 were shown to be highly corrupted with stripe artefacts, while non-problematic volumes were shown to have cSI lower than 1.15. For values between 1.15 and 1.25, some volumes were visually identified as corrupted by signal loss. This was observed only for volumes acquired along specific diffusion gradient directions, indicating that the dependency on the diffusion gradient is still present in the normalised parameter *cSI*. To remove these extra dependencies, a further correction is applied to the *cSI* values using the procedure illustrated in Fig. A.2. For all subjects, *cSI* values are organized in the order that the associated b-values and diffusion gradients directions were acquired (step 1 in Fig. A.2); the expected *cSI* profile along the diffusion gradient directions is computed from the mean values of *cSI* of non-problematic subjects, i.e. subjects that have any volume with *cSI* larger than 1.25 were removed (step 2 in Fig. A.2); all individual curves of *cSI* measures are then detrended by the expected *cSI* profile (step 3 in Fig. A.2). This procedure gives a value different from zero as its associated *cSI* value deviates from the expected *cSI* variance profile. Here, detrended "stripe" indexes will be referred to as the full corrected stripe index *fSI*. From visual inspections, this measure was able to distinguish volumes that were corrupted by motion artefacts from the ones that were not by setting a threshold value of 0.1.

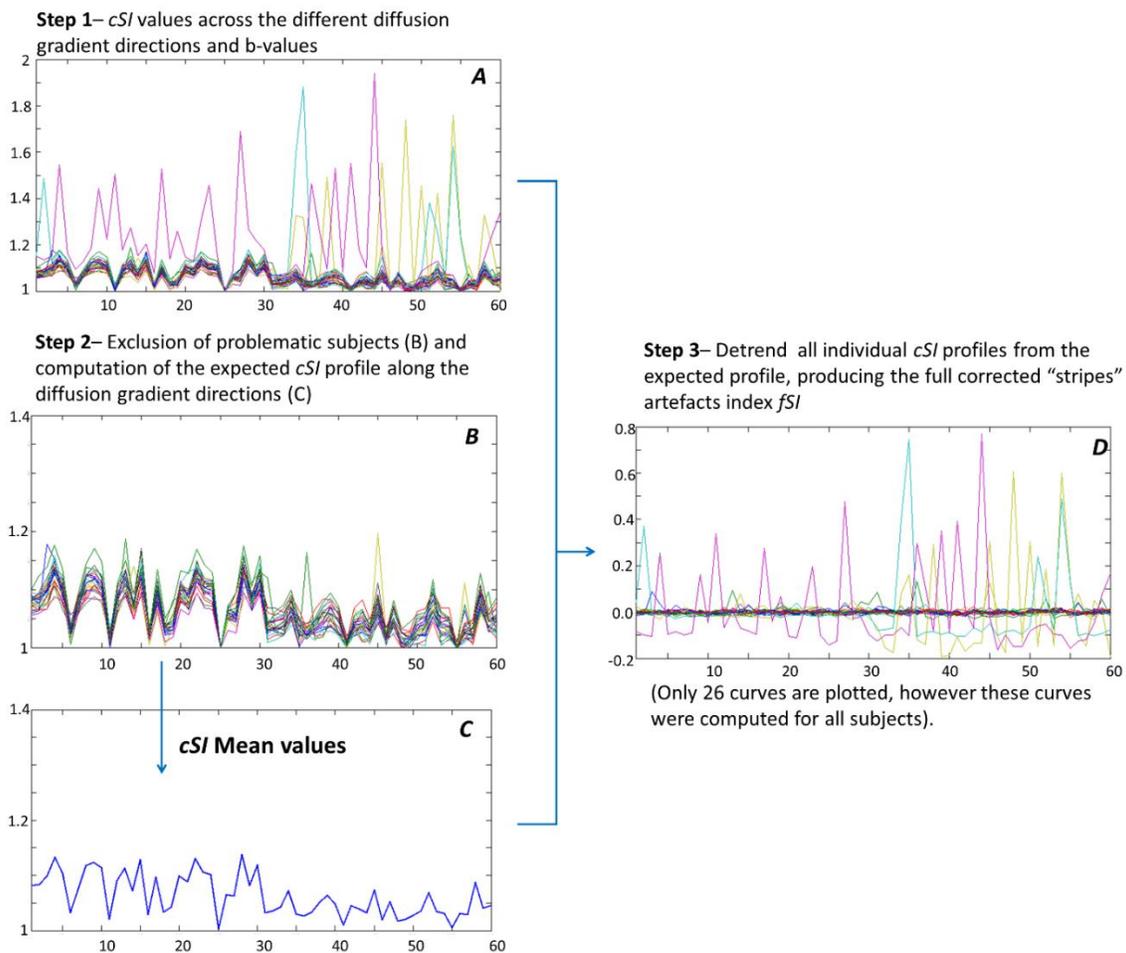

Fig. A.2 –Second phase of the procedure to quantify the artefacts induced by abrupt subject motion during data acquisition. (A) cSI values across the different diffusion gradient directions and b-values are plotted together. (B) cSI values for subject that showed a cSI value larger than 1.25 are removed; (C) cSI values for selected subjects are averaged; (D) cSI profiles from individual subjects are detrended from the averaged profiles obtaining the final fSI values.

## B. Correcting for motion-induced image misalignments

In this study, motion misalignments are corrected by registering each diffusion-weighted volume to a simulated template, allowing for the inherent contrast differences expected from dMRI acquisitions at different b-values and diffusion gradient directions.

The steps to generate the specific templates are summarized in Fig. B.1. Firstly, select the data acquired with b-value = 1000s/mm$^2$, avoiding larger amount of motion on the higher b-value data (step 1) to generate a first version of the templates. This first version is generated using the predicted signals of a diffusion tensor model fit to the selected b-value = 1000s/mm$^2$ data (step 2). To reduce the impact of possible artefacts on the low b-value regime, the data is smoothed (Gaussian kernel with FWHM = 1.25 × voxel size) and non-brain voxels removed before DTI fitting. The preliminary version of the templates is used to correct misalignments from the initial b-value data (step 3). The final version of the templates is then produced as the signal prediction of the diffusion tensor model re-fitted to the aligned data. These final versions of the templates are then produced for all b-values and gradient directions of the original dataset (step 4).

After the generation of templates, each original diffusion-weighted volume is aligned to its corresponding template using six degrees of freedom registration with the residual sum of squares minimization cost function. To preserve the correct association between the diffusion acquisition protocol and the image contrast, the motion parameters computed from each image registration are used to rotate the corresponding diffusion-weighted gradient direction.

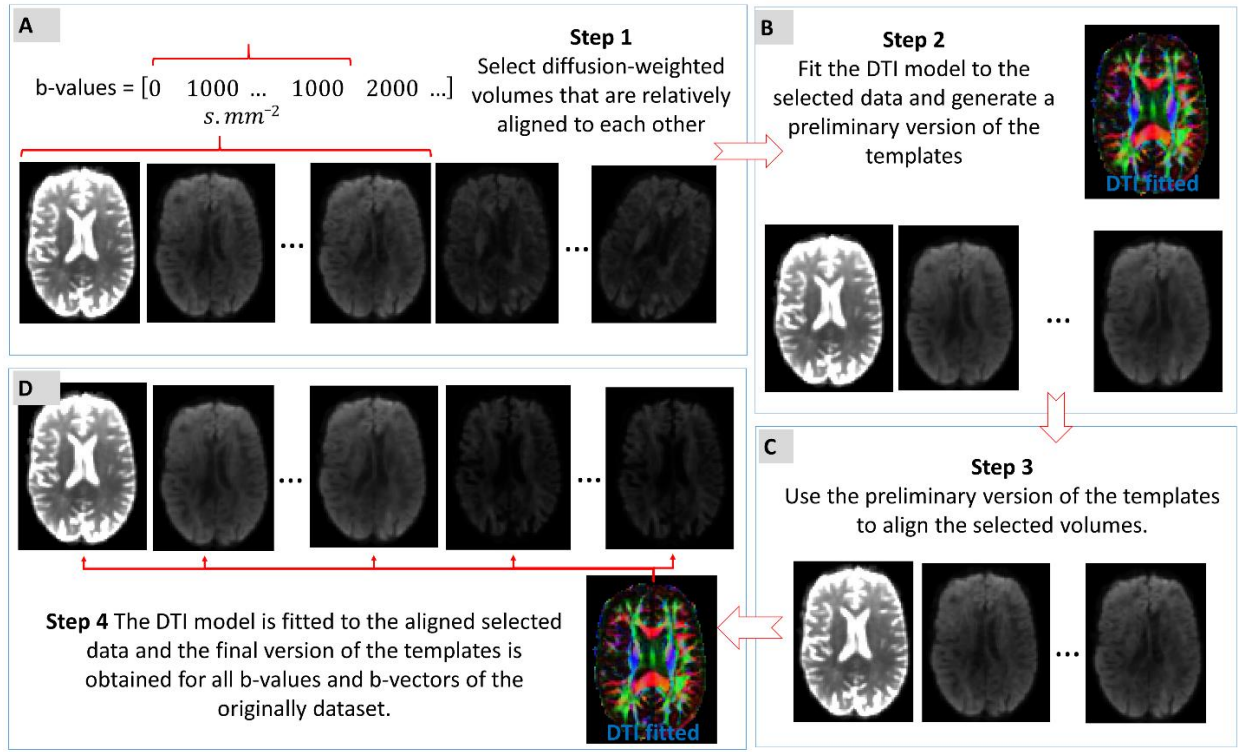

Fig. B.1 – Procedure to generate templates specific to the diffusion gradient directions and b-values used to acquire a diffusion-weighted dataset. A) Procedure step 1 - selection of the data for the lower b-value = 1000s/mm$^2$ which showed to be less corrupted by motion misalignments; B) Procedure step 2 – generation of the first version of data templates based on the selected lower b-value data; C) Procedure step 3 – realignment of the lower b-value data using the preliminary version of templates; D) Procedure step 4 – generation of the final version of data templates using the predicted signals for all original data b-values and diffusion gradient directions from the diffusion tensor model fitted to the lower b-value aligned data.